\documentclass{ws-ijmpa}
\usepackage[mathscr]{eucal} 
\usepackage{cite}

\usepackage{dcolumn}% Align table columns on decimal point 
\usepackage{bm}% bold math 
\newcommand{\be}{\begin{equation}} 
\newcommand{\ee}{\end{equation}} 
\newcommand{\bea}{\begin{eqnarray}} 
\newcommand{\eea}{\end{eqnarray}} 
\newcommand{\alp}{\alpha'}

\newcommand{\comment}[1]{}
\newcommand{\phis}{\phi_{\textrm{strg}}}
\newcommand{\psis}{\psi_{\textrm{strg}}}
\newcommand{\mpl}{m_{\textrm{Pl}}}

\newcommand{\gs}{g_{\textrm{s}}}

\newcommand{\eq}[2]{\begin{equation}\label{#1} #2 \end{equation}} 
\newcommand{\N}{\mathcal{N}}
\newcommand{\M}{\mathcal{M}}
\newcommand{\R}{\mathscr{R}}

\newcommand{\Teqpsi}{T_{\textrm{eq},\psi}}

\begin{document}

\date{\today} 
 
\title{Entropy Fluctuations in Brane Inflation Models}

\author{Robert H. Brandenberger}
\address{Department of Physics, McGill University, 3600 University Street\\
Montr\'eal, QC, H3A 2T8, Canada\\ email: rhb@hep.physics.mcgill.ca} 
\author{Andrew R. Frey }
\address{Department of Physics, McGill University, 3600 University Street\\
Montr\'eal, QC, H3A 2T8, Canada\\ email: frey@hep.physics.mcgill.ca} 
\author{Larissa C. Lorenz} 
\address{Institut d'Astrophysique de Paris, 98bis boulevard Arago\\ 75014
Paris, France\\ email: lorenz@iap.fr} 
 
\maketitle
 
\begin{abstract} 

We study the development of entropy fluctuations in brane inflation in
a warped throat, including the brane-antibrane tachyon as the waterfall
field. We find
that there is a period at the end of inflation during which the
entropy mode associated with the tachyon field increases exponentially.
In turn, the induced entropy seeds a contribution to the curvature
fluctuation on cosmological scales 
which grows rapidly and could exceed the primordial
curvature perturbation. We identify parameter values for which in the
absence of back-reaction the
induced curvature fluctuations are larger than the primordial
adiabatic ones. In the specific model we study, however, back-reaction limits
the growth of the entropy fluctuations. We discuss situations in
which back-reaction effects are less constraining.
The lesson of our investigation is that the study of the development 
of entropy fluctuations at the
end of the period of inflation can lead to constraints
on models of brane inflation 
and suggests that the curvaton mechanism may contribute significantly
to the spectrum of cosmological perturbations.

\end{abstract} 
 
\maketitle 
  
\section{Introduction} 

In recent years there has been a large effort at inflationary model
building in the context of superstring theory. Since string theory
contains many scalar field excitations and several of these
are massless above the scale of supersymmetry breaking, the hope
is that slow-roll inflation could be realized naturally in this
context (for recent reviews of inflation in the context of
string theory see \cite{hep-th/0501179, hep-th/0503195, 0708.2865, 0710.2951}).

A widely studied class of string-inspired inflationary universe models
falls under the category of brane inflation 
\cite{hep-ph/9812483,hep-th/0105032,hep-th/0105203,hep-th/0105204} (see
\cite{Burgess:2002kx,hep-th/0510018,hep-th/0610221} for reviews).
Here, the inflaton, the scalar field driving inflation, corresponds to
the separation between two branes, or between a brane and an antibrane
in a higher dimensional spacetime. When the branes come within a
critical distance (given by the string scale) from each other, a
tachyon develops (``tachyon condensation'') and inflation ends. The
dynamics of the inflaton in brane inflation is similar to the dynamics
of hybrid inflation \cite{astro-ph/9307002} where inflation ends when
a second scalar field, the ``waterfall'' field, develops a tachyonic
mass.

In this paper, we will focus on the dynamics of the initial stages of
reheating in brane inflation models. 
Specifically, we will study the growth of metric fluctuations of
entropic type on super-Hubble scales. We find that due to the
tachyonic growth in the entropy field, there is an instability to the
growth of metric entropy fluctuations on super-Hubble (but
sub-horizon) scales.  The entropy mode, in turn, induces a growing
curvature fluctuation which can dominate over the primordial
curvature perturbation for certain values of the model
parameters. Demanding that these induced curvature
fluctuations do not exceed the observational bounds leads to
constraints on the parameter space of brane inflation models. We begin
by setting the stage for our study.

In inflation models of chaotic type \cite{Linde:1983gd} (also called
large-field models), inflation ends when the inflaton field begins to
oscillate about the minimum of its potential energy function. As first
shown in \cite{Traschen:1990sw} (see also \cite{Dolgov:1989us}), these
oscillations lead to a parametric resonance instability for
fluctuation modes of the inflaton and of fields which couple to the
inflaton. This instability rapidly drains the energy from the
homogeneous inflaton condensate. This short initial stage of the
reheating process is called ``preheating'' \cite{hep-th/9405187}. As
shown in \cite{hep-ph/9407247}, the parametric resonance instability
persists even if the expansion of the universe is taken into account.
For a detailed discussion of preheating see \cite{hep-ph/9704452}.

As was discovered in \cite{hep-ph/0012142,hep-th/0106179} 
(see also \cite{hep-ph/9705357}),
the instability to the growth of field fluctuations is qualitatively
more efficient in models of hybrid inflation and is
called ``tachyonic preheating.'' Here, the
instability is fueled by the tachyonic mode of the background
model. In particular, in tachyonic preheating all long wavelength modes
are unstable.

Since the metric always couples to the inflaton, it is
not far-fetched to expect that oscillations of the inflaton
might induce instabilities of metric fluctuations. In fact, 
it was first suggested in \cite{hep-ph/9808404} that parametric excitation
of super-Hubble metric fluctuations during reheating might
be possible. However, in single field inflation models,
this effect does not occur \cite{hep-ph/9809490,hep-ph/9912510,gr-qc/0011075} 
because the
instability is in a gauge mode. However, the effect can be physical 
in a two field inflation model \cite{hep-ph/9901319}, and
concrete models were discussed in \cite{hep-ph/9909353,Fabio2}. In
\cite{Fabio2}, a specific hybrid inflation model was studied
as an example. In some of the models studied in \cite{Fabio2},
the metric fluctuations become nonlinear before back-reaction
can stop the instability \cite{hep-ph/0007219}.

In this paper, we will study the excitation of entropy modes of metric
fluctuations in brane inflation models of the type considered in
\cite{KKLMMT}.  Since the dynamics of preheating is of tachyonic type,
we find that there is a homogeneous solution for the entropy mode
which increases exponentially for a short time period at the end of
inflation.\footnote{In order for the instability to be excited, there
needs to be an offset of the tachyon field from its symmetric (and
unstable) point when averaged over a volume which corresponds to the
comoving Hubble volume at the onset of the period of inflation. A
similar assumption was also made in the analysis of \cite{Fabio2}.}
We show that this growth seeds an exponential instability of long
wavelength metric entropy modes, which turns off 
either once the tachyon field develops a sufficiently large velocity or
when back-reaction effects shut off the resonance.  We then
compute the magnitude of the induced curvature perturbation 
without taking into account back-reaction and find
that for certain parameter values it exceeds the amplitude of the
initial curvature fluctuation (one can
view this as a particular example of the curvaton scenario
\cite{Mollerach:1989hu,gr-qc/9502002,astro-ph/9610219,hep-ph/0110096,hep-ph/0110002,hep-ph/0109214}). 
A lesson we thus learn is
that the presence of entropy modes needs to be taken into account
in brane inflation models. However, in the specific model
we study, back-reaction effects may shut off the tachyonic resonance
of the fluctuations before the entropy modes reach a sufficient amplitude.

We should stress at the outset that we are only considering one
of several entropy modes present in our brane inflation model. We have
set all other modes (\textit{e.g.} modes due to motion in
angular directions of the compactification) to zero. An
interesting problem for further research would be to perform a
systematic study of all of the entropy modes which are present in the
setup.  In recent work \cite{Easson:2007dh,0709.3757,Huang:2007hh}, 
there has been
progress concerning multiple field perturbations (radial and angular
modes) while inflation is still under way.

There has been some previous work on the generation of secondary
fluctuations in brane inflation models. Closest to our work is
the study of \cite{Shandera} in which the secondary curvature
fluctuations due to fluctuations in the tachyon field were considered.
However, in that work the tachyonic amplification of the fluctuations
after the end of slow-roll inflation was not considered, and as a
consequence a much smaller amplitude of secondary perturbations was
found. The generation of isocurvature fluctuations at the end
of inflationary models of hybrid type due to inhomogeneities in
other light fields has been considered by various authors
\cite{Matsuda,Kolb}, as has ``modulated preheating" \cite{Dvali,Kofman},
\textit{i.e.},
the generation of inhomogeneities from variations of coupling constants
in the hybrid inflation model. In this case, these variations are due to
entropic fluctuations of other light fields which determine the
values of the coupling constants \cite{Lyth3,Lyth2,Lyth,Bernardeau}.
Some effects of higher order in perturbation theory have been
considered in   \cite{astro-ph/0601481,Barnaby}. Finally, we wish to draw the
attention of the reader to the interesting problem of the transfer
of energy from the inflaton/tachyon system to matter of the
Standard Model, the actual reheating process
\cite{BBC,Kofman2,Chialva,Leblond,Chen,FMM}.

The outline of this paper is as follows. In the following
section, we review the background geometry of warped brane
inflation models. Then we discuss the forms of the scalar
field potential which describe, respectively, 
the inflationary phase and the tachyon condensation period.
Section 4 contains a discussion of the background dynamics.
In section 5, we begin with a review of the formalism of
metric entropy fluctuations, which we proceed to apply
to our model. Finally, we study the growth of metric entropy perturbations in
our model and confront our conclusions with observational constaints.

\section{The background for warped brane inflation}

We take the basic setup of a brane moving in a warped throat, as 
in \cite{KKLMMT}, as our prototypical brane inflation model.  The most
important point of \cite{KKLMMT} is that the moduli of the compactification
manifold must be stabilized for inflation to occur, so the authors
of \cite{KKLMMT} focus on a class
of models in type IIB string theory in which fluxes fix the complex 
structure moduli   \cite{GKP,Keshav,Greene:2000gh,Kachru:2002he,Frey:2002hf} 
and a nonperturbative superpotential fixes the K\"ahler moduli \cite{KKLT}.
In these models, the internal geometry is (conformally) Calabi-Yau, which
can include singular points.  In the most studied case, the singularity
considered is the conifold, which may be deformed by a modulus to remove the
singularity.  In fact, the flux stabilizes the deformation modulus to a 
finite value, so the conifold becomes nonsingular.  In addition, the flux 
sources a warp factor, which causes the region near the deformed conifold
point to become a warped throat.   (For a review of these compactifications,
see for example \cite{Frey:2003tf}.)

It is also possible to introduce mobile D3 branes and antibranes into this
background.  Due to the warp factor, the antibranes sink to the bottom of
the throat (near the deformed conifold point) and contribute a positive 
supersymmetry breaking term to the vacuum energy \cite{KKLT}.  To a first
approximation, the D3 branes move without constraint, so they experience a
Coulomb attraction towards the antibranes, which (hopefully) drives inflation
\cite{KKLMMT}. 
The (more complicated) details of this situation are discussed below.

In the following, we first review the geometry of the background before and
after the flux deforms the conifold singularity. To conclude this section,
we then list the values of the key parameters which were assumed in
\cite{KKLMMT}.

\subsection{Singular Conifold}

A compact Calabi-Yau manifold may contain a variety of singularities, one
of which is known as the conifold, due to the fact that it is a
cone over the Einstein manifold $T^{1,1}$.  
Focusing on the region around the singularity, the Calabi-Yau metric is
approximated by the (noncompact) conifold metric
\eq{singular}{ds_{\textrm c}^2=dr^{2}+r^{2} ds_{T^{1,1}}^{2}\ .}
The base space $T^{1,1}$ has the metric
\cite{Candelas:1989js}
\begin{equation}\label{eq:T11}
ds_{T^{1,1}}^{2}=\frac{1}{9}\,(g_{5})^{2}+\frac{1}{6}
\sum_{i=1}^{4}(g_{i})^{2}\ ,
\end{equation}
the $g_{i}$ denoting a convenient basis of one-forms,
\begin{eqnarray}
g_{1}=\frac{e_{1}-e_{3}}{\sqrt{2}}&,&g_{2}=\frac{e_{2}-e_{4}}{\sqrt{2}}\ ,\\
g_{3}=\frac{e_{1}+e_{3}}{\sqrt{2}}&,&g_{4}=\frac{e_{2}+e_{4}}{\sqrt{2}}\ ,\\
&&g_{5}=e_{5}\ .
\end{eqnarray}
The $e_i$ are a vielbein
\begin{eqnarray}
e_{1}&=&-\sin\theta_{1} d\phi_{1}\ ,\\
e_{2}&=&\textrm d\theta_{1}\ ,\\
e_{3}&=&\cos\psi\,\sin\theta_{2} d\phi_{2}-\sin\psi d\theta_{2}\ ,\\
e_{4}&=&\sin\psi\,\sin\theta_{2} d\phi_{2}+\cos\psi d\theta_{2}\ ,\\
e_{5}&=&d\psi+\cos\theta_{1} d\phi_{1}+\cos\theta_{2} d\phi_{2}\ .
\end{eqnarray}

In type IIB string theory, the full 10D metric allowing for 4D 
spacetime-filling D3 branes at the tip of the conifold takes the form
\cite{hep-th/9807080} 
\begin{equation}\label{eq:simple}
ds^{2}=h^{-1/2}(r)G_{\mu\nu}dx^\mu dx^\nu+h^{1/2}(r)
ds_{\textrm c}^2\ .
\end{equation}
The external metric $G_{\mu\nu}$ is Minkowski for the known solutions of the
10D field equations, but we will allow it to take FRW form here.  Allowing
FRW evolution should introduce corrections to this ansatz, but we will assume,
as is standard, that they are small.

The warp factor $h(r)$ can be calculated from the 10D equations of motion to 
read
\begin{equation}\label{eq:h}
h(r)=1+\frac{R^{4}}{r^{4}}\ ,\ \ R^{4}=4\pi\gs\alpha'{}^{2}\frac{\N}{v}\ ,
\end{equation}
where $\N$ is the number of background D3 branes and $v$ is the volume ratio
$v=\textrm{Vol}\,T^{1,1}/\textrm{Vol}\, S^{5}$, \textit{i.e.} it
compares the size of the conifold base to a unit sphere.
As is standard, $\gs$ is the string coupling and $\alpha'$ is the
squared string length. Very near the tip of the conifold where $r$ is
small, $h\approx R^4/r^4$, and the 10D spacetime becomes
$\mathnormal{AdS}_5\times T^{1,1}$.  This region is the so-called
conifold throat, which joins on to the bulk Calabi-Yau at large $r$
when $h\approx 1$. Note that the warp fator (\ref{eq:h}) can be generalized
to a harmonic function on the conifold to account for more generally placed
D3 branes (or even wrapped D7 branes).  Henceforth, however, when we
use this form of the warp factor, we will assume that we can neglect the
constant term and that we have the simple form $h\approx R^4/r^4$.

\subsection{Deformed Conifold}
 
The singular conifold is a single point in the moduli space of conifold
metrics; it has been shown in \cite{GKP} that supergravity 3-form flux forces
the conifold onto the ``deformed conifold'' branch of this moduli space.
The deformed conifold asymptotes to the conifold, but it has a non-shrinking
3-cycle, so it is not a true cone near its core.  The deformed conifold
metric is \cite{Candelas:1989js}
\begin{eqnarray}
ds_{\textrm{dc}}^2&=&\frac{\epsilon^{4/3}}{2}\,K(\tau)\left\{\frac{1}{3K^{3}
(\tau)}\,\left[d\tau^{2}+(g_{5})^{2}\right]\right.\nonumber\\
&&\left.
+\cosh^{2}\left(\frac{\tau}{2}\right)\left[(g_{3})^{2}+(g_{4})^{2}\right]
\right.\nonumber \\
&&\left.
+\sinh^{2}\left(\frac{\tau}{2}\right)\left[(g_{1})^{2}+(g_{2})^{2}\right]
\vphantom{\frac{1}{K^3}}\right\}\ .
\label{eq:ds6warped}
\end{eqnarray}
These coordinates are dimensionless, so $\epsilon$ has dimensions of 
(length)$^{3/2}$.  In addition,
the function $K(\tau)$ in (\ref{eq:ds6warped}) is given by
\begin{equation}\label{eq:K}
K(\tau)=\frac{\left[\sinh(2\tau)-2\tau\right]^{1/3}}{2^{1/3}\sinh\tau}.
\end{equation}

In particular, the ``radial'' coordinate of the cone is now denoted by $\tau$ 
instead of $r$, as was the case for the singular conifold (\ref{eq:T11}). 
To convert between $\tau$ and $r$ in the asymptotic conifold region, 
note that $K(\tau)$ is asymptotically
\eq{kinfty}{
K(\tau\to\infty)\approx 2^{1/3}e^{-\tau/3}\ .}
Therefore, the radial and $g_5$ parts of the metric are
\begin{equation}
g_{\tau\tau}=g_{55}=\frac{\epsilon^{4/3}}{3\cdot 2^{5/3}}\,e^{2\tau/3}
\end{equation}
at large $\tau$.
Comparing this to the corresponding components of the singular conifold metric
(\ref{singular},\ref{eq:T11}) in the same 
limit, we find that $r$ and $\tau$ are related by
\bea
r^{2}&=&\frac{3}{2^{5/3}}\epsilon^{4/3}e^{2\tau/3}\ ,\nonumber\\
\tau&=&\frac{3}{2}\ln\left(\frac{2^{5/3}r^{2}}{3\epsilon^{4/3}}\right)
\equiv 3\ln\left(\frac{r}{r_0}\right)\ .\label{eq:r-and-tau}
\eea
Note that while $r$ has dimension of length, the new variable $\tau$ is 
dimensionless. This is a consequence of the new dimensional scale introduced 
into the theory by the deformation $\epsilon$. Also, $r_0$ is the naively
extrapolated value of the radius $r$ at the bottom of the throat.

In the presence of D3 branes (or D7 branes) or 3-form flux, the deformed
conifold develops a warp factor, as in the singular case, so the 10D metric
becomes \cite{hep-th/0007191} 
\begin{equation}\label{eq:ds2-warped}
ds^{2}=\hat{h}^{-1/2}(\tau) G_{\mu\nu}dx^\mu dx^\nu
+\hat{h}^{1/2}(\tau)\, ds_{\textrm{dc}}^{2}\ .
\end{equation}
(Regarding cosmological metrics, please see the comments following equation
(\ref{eq:simple}).)
The warp factor $\hat{h}(\tau)$ in the deformed case has a 
more complicated form, which is, for the case of no free D branes 
\cite{hep-th/0007191} (see also \cite{Herzog:2001xk} for more
explicit notation),
\begin{eqnarray}
\hat{h}(\tau)&=&\left(\gs\mathcal{M}\alpha'\right)^{2}2^{2/3}\epsilon^{-8/3}
\nonumber\\
&&\times\int_{\tau}^{\infty} dx\,\frac{x\coth x-1}{\sinh^{2}
x}\left[\sinh(2x)-2x\right]^{1/3} \ \ \label{eq:h-hat}
\end{eqnarray} 
up to an additive integration constant (usually taken to be 1 for a
throat attached to a compact Calabi-Yau manifold).  Free D3 branes
just add a harmonic piece to $\hat h$.  Physically, instead of
$\mathcal{N}$ D3 branes, we should now rather think of an
effective number of background branes
$\mathcal{N}_{\textrm{eff}}$: Apart from the $\mathcal{N}$ D3 branes,
the 3-form flux carries D3 brane charge (which is smeared over the
deformed conifold, rather than pointlike).

For large $\tau$, the
deformed conifold metric (\ref{eq:ds6warped}) leads back to
(\ref{singular},\ref{eq:T11}). The integral in (\ref{eq:h-hat}) cannot
be calculated analytically, but at large $\tau$ it is well
approximated by \cite{Herzog:2001xk,hep-th/0002159}
\begin{eqnarray}
\hat{h}(\tau)&\approx&\left(\gs\mathcal{M}\alpha'\right)^{2}3\cdot
2^{1/3}\epsilon^{-8/3}\tau e^{-4\tau/3} \nonumber \\
&=&\frac{81}{8}\frac{\left(\gs\mathcal{M} \alpha'\right)^{2}}{r^{4}}
\ln(r/r_{0})\label{happrox}
\end{eqnarray}
(again, up to an additive constant), where (\ref{eq:r-and-tau}) has
been used to obtain the last expression.  As we mentioned above, 
free D3 branes just add a harmonic term to (\ref{happrox}), which looks like
$R^4/r^4$ if the D3 branes are at the tip of the throat.
It is common to push the simplification still further and use the simple 
warp factor (\ref{eq:h}) even for the deformed conifold case, though this 
strictly speaking holds only over short distances.

\subsection{Warp factor at the bottom of the throat}

The presence of supergravity 3-form flux generates a potential for the
complex structure moduli, including the conifold deformation modulus
$\epsilon$.  In the approximation of a small deformation, 
$\epsilon^{2}$ is related to the flux by \cite{GKP}
\begin{equation}\label{eq:deformation}
\left(\epsilon^{2}\right)^{1/3}\approx\sqrt{\alpha'}\exp\left(-\frac{2\pi
\mathcal{K}}{3\mathcal{M}\gs}\right).
\end{equation}
Here, $\mathcal{M}$ and $\mathcal K$ are the quantum numbers of  
3-form flux wrapping different cycles. Their product gives the total 
effective 
D3 brane charge (modulo free branes) measured at the top of the throat, 
$\mathcal{N}=\mathcal{MK}$. We can integrate (\ref{eq:h-hat}) 
numerically to find that
\begin{eqnarray}
\hat{h}_{0}&=&
\hat{h}(\tau=0)=a\left(\gs\mathcal{M}\alpha'\right)^{2}\epsilon^{-8/3} 
\nonumber \\
&\approx& a\left(\gs\mathcal{M}\right)^{2}\exp\left(
\frac{8\pi\mathcal{K}}{3\mathcal{M}\gs}\right)\ ,\label{eq:hat-h0}
\end{eqnarray}
with $a\approx 1.14$.  Again, we assume that there are no free D3 branes in the
throat.

In fact, there is a simple physical argument due to \cite{GKP} that
the integral (\ref{eq:h-hat}) should be of order unity.  The warp
factor at a distance $r$ from $\N$ D3 branes is $h(r)\approx\gs \N
\alp{}^2/r^4$.  We have a natural distance scale $\epsilon^{2/3}$ at
the core of the deformed conifold, and (\ref{happrox}) tells us that
the effective number of D3 antibranes at $r\gtrsim r_0$ is $\gs\N\approx
(\gs\mathcal{M})^2$.  Therefore, we expect to have $\hat h_0\approx
(\gs\mathcal{M}\alp)^2\epsilon^{-8/3}$, which is indeed the case.

Just as we defined a naive value $r_0$ for the radius at the bottom of
the throat using (\ref{eq:r-and-tau}), we can define a similar value for the
radius by demanding that the simplest formula for the warp factor (\ref{eq:h})
(without the constant term)
gives the correct value (\ref{eq:hat-h0}) for the deformed conifold.  
Setting $\hat h_0\approx R^4/\hat r_0^4$, we find 
\eq{eq:hat-r0}{
\hat r_0 = \left(\frac{4\pi\gs \alp{}^2 \N}{v\hat h_0}\right)^{1/4} 
= \frac{2^{5/6}}{\sqrt{3}} \left(\frac{4\pi \mathcal{K}}{a \gs\mathcal{M} v}
\right)^{1/4} r_0\ .}

\subsection{Example values}\label{subsec:examples}

In the original KKLMMT paper \cite{KKLMMT}, the authors used 
the parameters (for the definition of the brane tension 
see (\ref{eq:T3}))
\begin{equation}
\frac{T_{3}}{\mpl^{4}} \approx 10^{-3}\ ,\ \ \gs= 0.1\ ,
\end{equation}
and hence 
\begin{equation}
\alpha'\mpl^{2} \approx 6.4\ ,
\end{equation}
which is a small hierarchy between the Planck and string scales.
In addition, \cite{KKLMMT} took sample values
\begin{equation}
\mathcal{K} = 8\ ,\ \ \mathcal{M} = 20,\,\textnormal{ therefore }\,
\N = 160\ .
\end{equation}
Finally, the volume ratio was taken to be $v = 1$,
which is equivalent to saying that the base of the 
conifold is simply the 5-sphere
itself. For the Einstein space $T^{1,1}$, as in
(\ref{eq:T11}), we have $v=16/27\approx\mathcal{O}(1)$. More generally,
however, one can think of $v$ as a free parameter describing the
string background geometry. While (\ref{eq:T11}) is the only
explicitly resolved example of a warped throat, it can be viewed as
one realization of a a class of backgrounds for which $v$ can vary
over a large range of values.  For our purposes, we will take 
\eq{vval}{v=16/27\ .}

If we calculate the warp factor $\hat{h}_{0}$ from
(\ref{eq:hat-h0}), the above values give
\begin{equation}
\hat{h}_{0} \approx 1.6\cdot 10^{15}.
\end{equation}
We pause here to note that the value of $2.6\cdot 10^{14}$ given in 
\cite{KKLMMT} does not precisely correspond to
their choice of discrete parameters; however, this value 
can be achieved with a very small fractional change of $\gs$.

\section{The potential for warped brane inflation}

Inflation occurs in the warped throat due to the attraction between D3 branes
and antibranes.  Due to the warp factor, D3 antibranes sink quickly to the
bottom of the throat, but D3 branes are only drawn to the bottom of the 
throat due to their interaction with the antibranes and to nonperturbative
effects related to moduli stabilization.  In this section, we briefly
review the potential that we use.

\subsection{Branes and antibranes}

As we mentioned before, the conifold and deformed conifold backgrounds 
allow for free, mobile D3 branes.  At the classical level, these D3 branes
and the background are mutually supersymmetric, so there is no force on
the brane.  D3 antibranes, however, feel a large classical force due to the
warp factor, which draws them to the bottom of the 
throat.\footnote{Technically, the antibranes cannot be introduced into the
compactification at the classical level \cite{GKP}, but we ignore that
complication as we introduce nonperturbative physics anyway.}  For technical
reasons, it is standard to assume that there is only a single antibrane at
the bottom of the throat.  Also at the
classical level, the D3 branes and antibranes experience a Coulombic 
attraction, which draws the D3 brane to the end of the throat.

The situation is more interesting when we consider nonperturbative effects.
First of all, the nonperturbative corrections to the potential stabilize
any moduli that remain unstabilized by the flux.  These terms also generate
a potential for D3 brane positions.  The effective 
4D potential derived from the 10D type IIB action is quite complicated when
all the moduli stabilization effects are correctly taken into account
\cite{hep-th/0607050}; for
the potential in several different cases, see  
\cite{arXiv:0705.3837,arXiv:0706.0360,arXiv:0705.4682,arXiv:0707.2848}.  
In the cases that
support slow-roll inflation (which requires some tuning), inflation occurs
near an inflection point far from the bottom of the throat.  Therefore,
by the time the brane reaches the bottom of the throat, it will be rolling
quickly.  However, \textit{since we will find that an increase in the
velocity of $\phi$ will only increase the production of entropy modes}, 
we use the naive Coulomb potential between the D3 branes and antibranes
to provide a lower limit.  In the absence of nonperturbative corrections,
this potential allows slow-roll inflation even when the D3 brane approaches
the bottom of the throat.

\subsection{Inflationary potential}
While inflation is under way in the throat (but far away from the 
bottom $r_{0}$), the inflaton field $\phi$ is just the  radial distance $r$ 
between the brane and the stack of anti-branes inside the throat,
normalized for a canonical kinetic term
\begin{equation}\label{eq:phi}
\phi=\sqrt{T_{3}}\,r\ ,
\end{equation} 
where $T_{3}$ is the D3 brane tension,
\begin{equation}\label{eq:T3}
T_{3}=\frac{1}{(2\pi)^{3}g_{s}\alpha'^{2}}\ .
\end{equation}
Note that $\phi$ has dimension of mass. 

From \cite{KKLMMT}, we have the Coulomb potential
\eq{eq:pot-inf}{
V^{\textrm{inf}}(\phi) = \frac{M^{4}}{1+\left(\frac{\mu}{\phi}\right)^{4}}
\approx M^{4}\left[1-\left(\frac{\mu}{\phi}\right)^{4}\right]\ ,
}
for $\phi\gg\mu$.  The two parameters $M$ (the
overall scale of inflation) and $\mu$ (scale compared to the field
value $\phi$ at any given moment) are related to the fundamental
string geometry parameters through 
\eq{eq:mandmu}{
M^{4} = \frac{4\pi^{2} v \hat\phi_{0}^{4}}{\N}\ ,\ \
\mu^{4} = \frac{\hat\phi_{0}^{4}}{\N} = \frac{M^{4}}{4\pi^{2}v}\ ,
}
where $\hat\phi_{0}=\sqrt{T_{3}}\,\hat r_{0}$ is the field value at the bottom
of the throat with $\hat r_{0}$ from (\ref{eq:hat-r0}).  (We use $\hat r_0$ 
rather than $r_0$ because the Coulomb potential arises from perturbations in
the warp factor.) Note in particular that
\begin{equation}\label{eq:v}
\frac{M}{\mu} = \left(4\pi^{2}v\right)^{1/4},
\end{equation}
so we can consider $\mu\approx M$ for $v\approx \mathcal{O}(1)$.

The assumptions made in deriving the potential 
(\ref{eq:pot-inf})\footnote{Which are that the brane and antibrane
interact by closed string modes and that the throat is well approximated
by equations (\ref{eq:simple},\ref{eq:h}).}
break down when the \emph{proper} distance between the
branes reaches the string scale.  We will discuss that situation in the
following.

Provided that the linear inflaton fluctuations seed the observed structure
in the universe, the mass scale $M$ of inflation can be determined
from the COBE constraint \cite{COBE}
\be \label{primcond}
\frac{\delta \rho}{\rho} \approx 
\frac{H \left(V^{\textrm{inf}}\right)'}{\dot \phi^2} \approx
10^{-5} \ ,
\ee
where a prime indicates a derivative with respect to $\phi$. Inserting
the potential (\ref{eq:pot-inf}), we obtain 
\bea
\left(\frac{M}{\mpl}\right)^3 &=&\frac{v^{1/4}}{4\sqrt{3}\pi} 
\frac{\delta\rho}{\rho} \left(\frac{\mu}{\phi_H}\right)^5\nonumber\\
&\approx& 4.4\cdot 10^{-8} \left(\frac{\mu}{\phi_H}
\right)^{5}\ ,
\eea
where $\phi_H$ is the value of the inflaton when scales of
cosmological interest exit the Hubble radius. $\phi_H$ is 
slightly larger than the value at the waterfall point, which we find
in equation (\ref{eq:phistrings}) below. If we use the waterfall value for
the parameter values of \cite{KKLMMT} (which is $\approx 25.3 \mu$), we find
\be \label{COBEvalue}
\frac{M}{\mpl} \approx 3.4\cdot 10^{-5} \ .
\ee
This value is consistent within about a percent
with the fact that the scale of the potential is
given by the D3 brane tension, $M^4=2\hat h_0^{-1}T_3$.

\subsection{Tachyonic potential}

We will now discuss what happens when the D3 brane approaches within a 
string length of the antibrane and simultaneously reaches the bottom of the
throat.

Since the brane is close to the bottom of the throat, we 
have to take into account the deformation of the conifold. In particular, $r$ 
ceases to be the appropriate radial coordinate. Near the bottom of
the throat, which we will take to be within a string length of the antibrane, 
the appropriate canonically normalized field is
$\psi$ defined by
\begin{eqnarray}
\psi &=& \sqrt{T_{3}}\,\frac{\epsilon^{2/3}}{3^{1/6}\cdot 2^{5/6}}\,\tau 
\nonumber\\
&=&\left(\frac{3}{2}\right)^{5/6} \sqrt{T_{3}}\,\epsilon^{2/3}\ln(r/r_{0})\ .
\label{eq:phihat}
\end{eqnarray}
Due to the presence of the factor $\epsilon^{2/3}$ in this 
rescaling, $\psi$ acquires the correct dimension of mass despite 
$\tau$ being dimensionless. 

Simultaneously the lightest
excitation of the open string stretching from brane to antibrane becomes
tachyonic. This
tachyon starts rolling down its potential, leading to tachyonic
reheating \cite{hep-ph/0012142,hep-th/0106179}. The total two-field potential
after the appearance of the tachyon $T$ (which replaces (\ref{eq:pot-inf}))
can be modeled as 
\begin{equation}\label{eq:pot-reh}
V^{\textrm{reh}}(\psi,T)=v_{0}^{4}+\hat{h}^{-1/2}_{0}\,\left\{
-\frac{1}{\alpha'}+\frac{\hat{h}_{0}^{1/2}\,\psi^{2}}{T_{3}
\left(2\pi\alpha'\right)^{2}}\right\}T^{2}
\end{equation}
(see, for example, \cite{hep-th/0303057,hep-th/0411222,0710.5469,
Polchinski:1998rq,Polchinski:1998rr}), which is a
valid approximation for small tachyon values.  Here,
$\hat{h}_{0}$ is the warp factor at the bottom of the throat from
(\ref{eq:hat-h0}).  
For simplicity, we have restricted the tachyon to
real values, but our argument is not affected by taking $T$ complex.

The potential (\ref{eq:pot-reh}) is reminiscent of that of hybrid inflation, 
where the role of the waterfall field being played by $T$. The 
``waterfall point'' occurs at a value $\psi=\psis$ which corresponds to a 
brane/antibrane separation of the string length:
\be
\psis=2\pi \sqrt{\alp T_3}\hat h_0^{-1/4} =2\pi a^{-1/4}
\sqrt{\frac{T_3}{\gs\mathcal{M}}}\epsilon^{2/3}\ ,
\label{eq:psistrings}
\ee
where we have used (\ref{eq:hat-h0}) in the last equality.  We can convert 
this value to the ``long-distance'' canonically normalized scalar
using (\ref{eq:hat-r0},\ref{eq:mandmu},\ref{eq:phihat})
\bea
\phis 
&=&\phi_0 \exp\left[ \frac{2\pi}{a^{1/4}\sqrt{\gs \mathcal{M}}}
\left(\frac{2}{3}\right)^{5/6}\right]\nonumber\\
&=&\mu \frac{\sqrt{3}}{2^{5/6}} \left[\frac{a\gs\M^2v}{4\pi}\right]^{1/4}
\exp\left[ \frac{2\pi}{a^{1/4}\sqrt{\gs \mathcal{M}}}
\left(\frac{2}{3}\right)^{5/6}\right]\nonumber\\
&\approx& 25.3\mu \ ,\label{eq:phistrings}
\eea
where the last approximation is for the example values from section
\ref{subsec:examples}. 
For $\psi\lesssim \psis$ (or equivalently $\phi\lesssim\phis$), 
the tachyon starts rolling down the inverted square potential from $T=0$.

Setting the potentials (\ref{eq:pot-inf}) and
(\ref{eq:pot-reh}) equal at $\phi=\phis$ and $T=0$ determines the
scale $v_0$:
\begin{equation}
v_0^{4}\approx M^{4} \, 
\end{equation}

\section{Background evolution}

Recall that during inflation, \textit{i.e.} while the potential is
given by (\ref{eq:pot-inf}), the inflaton rolls according to
\begin{equation}\label{eq:eofm-inf}
\ddot{\phi}+3H\dot{\phi}+V^{\textrm{inf}}_{\phi}  = 0 \ .
\end{equation}
Once preheating and therefore the potential (\ref{eq:pot-reh}) sets in, 
the equations governing the background fields' evolution are
\begin{eqnarray}
\ddot{\psi}+3H\dot{\psi}+V^{\textrm{reh}}_{\psi}&=&0\ ,\label{eq:eofm-phi}\\
\ddot{T}+3H\dot{T}+V^{\textrm{reh}}_{T}&=&0\ ,\label{eq:eofm-T}\\
\frac{8\pi}{3\mpl^2}\,
\left[\frac{\dot{\psi}^{2}}{2}+\frac{\dot{T}^{2}}{2}+V^{\textrm{reh}}
(\psi,T)\right]&=&H^{2}\ .\label{eq:Friedmann}
\end{eqnarray}

The astute reader might wonder whether we need to account for the fact that
the inflaton and tachyon have noncanonical kinetic terms in the proposed
D-brane action with the tachyon (see, \textit{e.g.},  
\cite{hep-th/0303057,hep-th/0411222,0710.5469,hep-th/0211180,hep-th/0003221}).
(Reheating with the full tachyon action, minus the inflaton, was 
studied in \cite{hep-th/0412095}.)  However,
as we will see, the velocities and field values we study are all smaller
than the local string scale, so we need not concern ourselves with this
issue.    In this, we agree with \cite{Barnaby}.

\subsection{Tachyon evolution}

Let us first consider the evolution of the tachyon during reheating.
Later we will find that entropy fluctuations grow when
$|\dot\psi|>|\dot T|$, so we are interested in finding the maximum
possible speed of the tachyon in order to set a lower limit on the
entropy perturbations.  Since we want the greatest tachyon derivative,
we can drop the $\psi^2 T^2$ term in the potential (\ref{eq:pot-reh}),
which only reduces the tachyonic mass.

Shortly after the tachyon has started rolling, its velocity $\dot{T}$
is still small, but the acceleration $\ddot{T}$ is large. Then the equation of
motion (\ref{eq:eofm-T}) can be approximated by
\begin{equation}
\ddot{T}\approx \frac{2}{\alpha'}\,\hat{h}^{-1/2}_{0}\,T\,.
\end{equation}
(in addition, this approximation overestimates the acceleration). 
Hence, the evolution of the tachyon is described by
\begin{equation}\label{eq:solT}
T(t)\approx T_{0}\exp\left[\left(\frac{\mathcal{C}_{1}}{\sqrt{\alpha'}}
\right)\,t\right],
\end{equation}
where $\mathcal{C}_{1}=\sqrt{2}\,\hat{h}_{0}^{-1/4}$. From this solution, we 
find that the velocity $\dot{T}$ is proportional to $T$ with
\begin{equation}\label{eq:Tdot}
\dot{T}\approx\left(\frac{\mathcal{C}_{1}}{\sqrt{\alp}}\right)\,T\ .
\end{equation}

We can therefore compare the time scale of the tachyon decay to the time 
scale of inflation, which from (\ref{eq:pot-inf}) and the Friedmann 
equation is given by $H\approx \left(M^{2}/\mpl\right)$. 
For the tachyon decay to be faster than the time scale of inflation, we
require
\begin{equation}
\left(\frac{\mathcal{C}_{1}}{\sqrt{\alp}}\right)^{-1}
\approx \frac{1}{\sqrt{2}}\,\hat{h}_{0}^{1/4}\sqrt{\alp}
< \frac{\mpl}{M^2}\ .
\end{equation}
Using the values of section \ref{subsec:examples}, \textit{i.e.},
$\hat{h}_{0}\approx 10^{15}$, and the value of $M/\mpl\approx 10^{-5}$ given
in (\ref{COBEvalue}), we find indeed that tachyon decay is faster than
the Hubble scale. This justifies, in retrospect, neglecting the Hubble
friction term in (\ref{eq:eofm-T}).

\subsection{Inflaton evolution}

We now proceed to determining the inflaton velocity at the beginning of 
reheating. Since $\phi$ (or the new field $\psi$, respectively) does not 
accelerate further once the potential shifts from (\ref{eq:pot-inf}) 
to (\ref{eq:pot-reh}) ($T$ is small initially), a good 
estimate for the energy in the $\phi$ (or $\psi$) field early on 
in the reheating 
phase can be obtained from  $\dot{\phi}_{\textrm{strg}}$, the inflaton 
velocity at the waterfall point $\phis$. To determine 
$\dot{\phi}_{\textrm{strg}}$, let us assume that the slow-roll approximation 
for $\phi$ is still valid when the field encounters $\phis$ moving down the 
potential (\ref{eq:pot-inf}). We remind the reader again that a large value
of $\dot\phi$ (or equivalently $\dot\psi$) will increase the amount of
entropy perturbations, so we are really finding a lower limit.

Simplifying (\ref{eq:eofm-inf}) 
to $3H\dot{\phi}\approx -V^{\textrm{inf}}_{\phi}$, 
we can estimate the velocity $\dot{\phi}_{\textrm{strg}}$ at the waterfall 
point:
\bea
\dot{\phi}_{\textrm{strg}}&=&-\sqrt{\frac{2}{3\pi}}\,\left(\frac{\mu}{\phis}
\right)^{4}\frac{M^2}{\phis}\,\mpl \nonumber\\
&=& -\frac{2}{\sqrt{3}} v^{1/4} M\mpl \left(\frac{\mu}{\phis}\right)^5\ .
\label{eq:phidot}
\eea

Note, however, that $\phi$ is the ``old'' field used during inflation,
and that from the appearance of the tachyon on, we must use
$\psi$ as defined in (\ref{eq:phihat}).  From this transformation, we
see that the velocities $\dot{\psi}$ and $\dot{\phi}$ are related by
\begin{equation}
\dot{\psi}=\left(\frac{3}{2}\right)^{5/6}\sqrt{T_{3}}\,\epsilon^{2/3}\,\frac{
\dot{\phi}}{\phi}\ , 
\end{equation}
so a velocity (\ref{eq:phidot}) in $\phi$ becomes a velocity for $\psi$:
\begin{equation}\label{eq:psidot}
\dot{\psi}_{\textrm{strg}}=-\left(\frac{3}{2}\right)^{1/3}
\sqrt{\frac{T_{3}}{\pi}}\,\epsilon^{2/3}\,\frac{\mu^{4}M^{2}}{\phis^{6}}\,
\mpl \ .
\end{equation}
(The astute reader should note that this velocity is, in fact, considerably
less than the local string scale.)
Therefore, the value of $T$ for which $|\dot{T}|>|\dot{\psi}|$ is
\begin{equation}\label{eq:hatTequal}
\frac{T}{\mpl}>\left(\frac{\sqrt{\alp}}{\mathcal{C}_{1}}\right)
\left(\frac{3}{2}\right)^{1/3}
\sqrt{\frac{T_{3}}{\pi}}\,\epsilon^{2/3}\,\frac{\mu^{4}M^{2}}{\phis^{6}}
\equiv\frac{T_{\textrm{eq},\psi}}{\mpl}.
\end{equation}
Simplifying a bit and inserting the values from section
\ref{subsec:examples}, we find 
\bea
\frac{T_{\textrm{eq},\psi}}{\mpl} &=&\frac{a^{1/4}}{2\pi}
\left(\frac{3}{2}\right)^{1/3}
\sqrt{v \mathcal{M}}
\left(\frac{\mu}{\phis}\right)^6 \nonumber\\
&\approx& 2.5\cdot 10^{-9} \, .
\eea
This value is in fact close to the Hubble scale and therefore 
much smaller than the local (warped) string scale
at the bottom of the throat, which is given by $M$.  Since the value of
the tachyon also controls the tachyon velocity, we see that the tachyon
velocity is also much smaller than the local string scale.

We now need to compare the value of $T_{\textrm{eq}, \psi}$ with the
initial value of the tachyon once the tachyonic instability sets in.
Note from the form of the potential (\ref{eq:pot-reh}) that the
tachyonic instability sets in rather suddenly as the inflaton rolls
down its potential. At large inflaton values, the tachyon is confined
in the steep valley and displaced from $T = 0$ by quantum
fluctuations.
An upper bound on the amplitude of the quasi-homogeneous
tachyon value is obtained by assuming that the mode carries
the typical energy density of quantum vacuum fluctuations 
\begin{equation}
\rho_{\mathrm{vac}} \approx H^4 \ ,
\end{equation}
which holds even for very
large masses ($m\gg H$) up to numerical coefficients of order $10^{-2}$
\cite{VF}.
The average dispacement $T_{0}$ of the tachyon can then be
estimated by equating the energy density in the tachyon displacement
with the above vacuum fluctuation energy density
\begin{equation} \label{amp}
m^2 T_{0}^2 \approx H^4 \ ,
\end{equation}
where $m$ is the local tachyon mass which can be read off from
(\ref{eq:pot-reh}).  
Furthermore, by examining the tachyon's positive frequency modes,
it is straightforward to see that the super-Hubble modes account for the
entire background. 
\emph{Within a Hubble patch}, one may therefore think of $T_{0}$ as a
homogeneous background for the tachyon; while there are fluctuations on shorter
wavelengths, they decouple at linear order and affect only the
backreaction, as we will discuss later.

However, a refined argument shows that 
$T_0$ as given in (\ref{amp}) is not truly a homogeneous
background for the tachyon.  Because the tachyon is massive compared to the
Hubble scale during most of inflation, tachyon fluctuations at large 
wavelength are suppressed by the expansion of the universe.
Therefore, 
if we are interested in studying fluctuations on a particular
super-Hubble scale $k$, we need to use the \emph{quasi-homogeneous} part of
the tachyon field, not over a Hubble volume, but a much larger volume
of comoving radius $k^{-1}$. Since the tachyon field is massive during
inflation, the super-Hubble fluctuations are damped by $\exp(-2 N_k)$,
where $N_k$ is the number of e-foldings of inflation since the mode labelled
by $k$ exited the Hubble radius. This leads to an exponential suppression
of $T_{0}(k)$. To see the exponential suppression, we note that the
tachyon fluctuation at a specific wavenumber is 
\eq{Tatk}{\delta T(k)\sim \sqrt{k}/ma^2\ ,}
so the total integrated (root-mean-square)
background is 
\eq{Tamp}{T_0(k)\sim (1/m)(k/a)^2 = \exp(-2N_k)H^2/m\ .}
We will return to this point in section \ref{s:backreaction}.\footnote{We 
thank Jim Cline and Neil Barnaby for emphasizing this point to us.}
It is important to remember that $T_0$ is not a homogeneous value;
rather it is a \emph{quasi-homogeneous} 
background and does depend on wave number.
However, though the quasi-homogeneous background depends on the wavenumber,
it is justified to describe it as a $k$-\emph{independent, homogeneous}
background on the scales of interest.
In addition, we note that $T_0$ is considerably smaller than the Hubble scale,
so it is also smaller than $\Teqpsi$, which allows tachyonic growth.

\section{Entropy fluctuations}

In our model, there are two scalar fields at play during the phase of
reheating, namely the (coordinate-adjusted) inflaton $\psi$ and the
tachyon $T$. Therefore, as in any multifield inflation model, entropy
perturbations should be present.  If their growth is fast enough,
these entropy perturbations can act as a source for the comoving
curvature perturbation $\mathscr{R}$, which is conserved on large
scales in single field inflation. The change in $\mathscr{R}$ is
described by \cite{Gordon:2000hv}
\begin{equation}\label{eq:changeinR}
\dot{\mathscr{R}}=\frac{H}{\dot{H}}\,\frac{k^{2}}{a^{2}}\,\Psi
-\frac{2H}{\dot{\sigma}^{2}}\,V_{s}\,\delta s\ .
\end{equation}
In the above, $\Psi$ is the metric fluctuation in longitudinal
gauge, in the gauge in which the metric including linear cosmological
perturbations (in the absence of anisotropic stress)
takes the form
\begin{equation}
ds^2 \, = \, (1 + 2 \Psi) dt^2 - a^2(t)(1 - 2 \Psi) d\vec{x}^2 \,
\end{equation}
and $\delta s$ is the entropy perturbation. See \textit{e.g.}
\cite{MFB} for an in-depth survey of the theory of cosmological
perturbations, and \cite{RHBrev} for a pedagogical overview.

Moreover, $\dot{\sigma}$ is the ``adiabatic'' combination of the 
field derivatives $\dot{\psi}$ and $\dot{T}$, defined as\footnote{For the 
generalization of these expressions to more than two fields, 
see \cite{Wands:2007bd}.}
\begin{eqnarray}
\dot{\sigma} &=& \sqrt{\dot{T}^{2}+\dot{\psi}^{2}}\nonumber\\
&=&\cos(\theta)\,\dot{\psi} + \sin(\theta)\,\dot{T}\label{eq:sigmadot}
\end{eqnarray}
with
\begin{equation}
\cos(\theta) = \frac{\dot{\psi}}{\sqrt{\dot{\psi}^2 + \dot{T}^2}} \ ,\ \
\sin(\theta) = \frac{\dot{T}}{\sqrt{\dot{\psi}^2 + \dot{T}^2}} \ .
\end{equation}
Orthogonal to the adiabatic direction 
$\sigma$  in field space, the ``entropy direction'' $s$ is given by
\begin{equation}
\delta s = \cos(\theta) \,\delta T - \sin(\theta) \,\delta \psi \ .
\end{equation}
Note that the background entropy field
is constant, $\dot{s}=0$, and
can therefore be set to zero. Finally, $V_{s}$, the potential's derivative in
the entropy direction, can be expressed as the following combination
of the potential derivatives:
\begin{equation}\label{eq:V_s}
V_{s}=\frac{\dot{\psi}}{\dot{\sigma}}\,V_{T}-\frac{\dot{T}}{\dot{\sigma}}\,
V_{\psi}
\end{equation}
Note that we have dropped the index ``reh'' from the potential. 
Unless otherwise specified, $V$ from now on always refers to the 
two-field potential (\ref{eq:pot-reh}).

The evolution equation 
for the entropy perturbation is
\begin{eqnarray}\label{eq:eofm-ds}
\delta\ddot{s} &+& 3H\delta\dot{s} + 
\left(\frac{k^{2}}{a^{2}}+V_{ss}+3\frac{V_{s}^{2}}{\dot{\sigma}^{2}}
\right)\delta s \nonumber \\
&=& \frac{\dot{\theta}}{\dot{\sigma}}\,\frac{k^{2}}{2\pi G a^{2}}\,\Psi\,,
\end{eqnarray}
which contains also the second derivative of the potential with respect to 
the entropy field, given by
\begin{equation}\label{eq:V_ss}
V_{ss}=\frac{\dot{T}^{2}}{\dot{\sigma}^{2}}\,V_{\psi\psi}-2\,\frac{\dot{\psi}
\dot{T}}{\dot{\sigma}^{2}}\,V_{\psi T}+
\frac{\dot{\psi}^{2}}{\dot{\sigma}^{2}}\,V_{TT}\ .
\end{equation}

Evidently, the source term on the right hand side of
(\ref{eq:eofm-ds}) dies out at large scales, $k/a\rightarrow 0$, as
does the first term on the right hand side of
(\ref{eq:changeinR}). Therefore, only the term proportional to $\delta
s$ remains in (\ref{eq:changeinR}) as a source for $\mathscr{R}$ on
large scales. To understand its evolution, we need to solve the
equation resulting from (\ref{eq:eofm-ds}) in the limit
$k/a\rightarrow 0$,
\begin{equation}
\delta\ddot{s}+3H\delta\dot{s}+\left(V_{ss}+3\frac{V_{s}^{2}}{\dot{\sigma}^{2}}
\right)\delta s\approx 0\ .
\end{equation}

First, we calculate explicitly the various derivatives of the potential,
\begin{eqnarray}
V_{\psi} &=& \frac{2\psi}{T_{3}}\,\frac{T^{2}}{(2\pi\alpha')^{2}} 
= 4\pi\gs T^{2}\psi\ , \nonumber \\
V_{T} &=& 2
\hat{h}^{-1/2}_{0}T\,\left(-\frac{1}{\alpha'}+
\frac{\hat{h}_{0}^{1/2}\psi^{2}}{T_{3}(2\pi\alpha')^{2}}\right) \nonumber\\
V_{\psi\psi} &=& \frac{2}{T_{3}}\,\frac{T^{2}}{(2\pi\alpha')^{2}}
= 4\pi\gs T^{2}\ , \nonumber\\ 
V_{TT} &=& 2
\hat{h}^{-1/2}_{0}\,\left(-\frac{1}{\alpha'}+
\frac{\hat{h}_{0}^{1/2}\psi^{2}}{T_{3}(2\pi\alpha')^{2}}\right)\ , \nonumber \\
V_{\psi T} &=& \frac{2\psi}{T_{3}}\,\frac{2T}{(2\pi\alpha')^{2}}=8\pi\gs T\psi
\ ,\label{eq:potderivs}
\end{eqnarray}
where occasionally (\ref{eq:T3}) has been used to replace $T_{3}$.

The entropy mode $\delta s$ grows exponentially if the ``mass'' term in 
(\ref{eq:eofm-ds}) is negative. Of the two terms 
$V_{ss}+3V_{s}^{2}/\dot{\sigma}^{2}$, only the first one can be negative, 
so we need 
\begin{equation}\label{eq:condition}
\frac{|V_{ss}|}{3V_{s}^{2}/\dot{\sigma}^{2}}>1\ , \ \ V_{ss}<0
\end{equation}
to obtain a tachyonic mass for $\delta s$. The evolution of the
background fields discussed in the previous section now allows us to
identify the dominant terms in $V_{ss}$ and
$V_{s}^{2}/\dot{\sigma}^{2}$.

\section{Growth of the entropy fluctuations}

Once the D3 brane has come within a string length of
the antibrane, there occurs
a short period $0<T<T_{\textrm{eq},\psi}$, during which $\dot{T}$ is
catching up with $\dot{\psi}$. After that, for
$T>T_{\textrm{eq},\psi}$, the fields definitely start to roll down in
the $T$-direction in field space. Let us reexamine the condition
(\ref{eq:condition}) for tachyonic increase of the entropy mode in
the light of this fact. 
For the moment, we neglect the effects of back-reaction.

\subsection{Limit $|\dot{\psi}|>|\dot{T}|$}

In this limit, we see from (\ref{eq:sigmadot}) that we have 
$\dot{\sigma}\approx \dot{\psi}$. Hence, equations (\ref{eq:V_s}) and 
(\ref{eq:V_ss}) reduce to
\begin{eqnarray}
V_{ss}&=&\frac{\dot{T}^{2}}{\dot{\psi}^{2}}\,V_{\psi\psi}-2\,
\frac{\dot{T}}{\dot{\psi}}\,V_{\psi T}+V_{TT}\approx V_{TT}<0\ ,\\
V_{s}&=&V_{T}-\frac{\dot{T}}{\dot{\psi}}\,V_{\psi}\approx V_{T}\ .
\end{eqnarray}
Using the explicit expressions (\ref{eq:potderivs}) for the
derivatives, we have
\begin{equation}
\frac{|V_{ss}|}{3V_{s}^{2}/\dot{\sigma}^{2}}
\approx \frac{|V_{TT}|\dot{\psi}^{2}}{3V_{T}^{2}}
\approx \frac{\dot{\psi}^{2}\alpha'}{3\mathcal{C}_{1}^{2}T^{2}}\ .
\end{equation}
But we see by the definition of $\Teqpsi$ (assuming 
$\dot\psi=\dot\psi_{\mathrm{strg}}$ constant) that 
\eq{Vss}{\frac{|V_{ss}|}{3V_s^2/\dot\sigma^2} \approx \frac{\Teqpsi^2}{3T^2}
\ .}
 
For small values of $T$, this ratio is much larger than unity, and
hence the condition for tachyonic instability of the entropy mode
is satisfied. In fact, the tachyonic instability for the entropy mode
ends just before $|\dot T|=|\dot\psi|$.  Thus, the tachyonic
resonance of the entropy mode continues up to the time when the
tachyon velocity starts to exceed the inflation velocity.

\subsection{Limit $|\dot{T}|>|\dot{\psi}|$}

In this limit, equation (\ref{eq:sigmadot}) gives
$\dot{\sigma}\approx\dot{T}$, and therefore from (\ref{eq:V_s}) and
(\ref{eq:V_ss}) one finds
\begin{eqnarray}
V_{ss}&=&V_{\psi\psi}-2\,\frac{\dot{\psi}}{\dot{T}}\,V_{\psi
T}+\frac{\dot{\psi}^{2}}{\dot{T}^{2}}V_{TT}\approx V_{\psi\psi}\ ,\\
V_{s}&=&\frac{\dot{\psi}}{\dot{T}}V_{T}-V_{\psi}\approx -V_{\psi}\ .
\end{eqnarray}
Since $V_{\psi\psi}$ is positive, we see that there is no
tachyonic resonance in this region. Thus, in order to have any
tachyonic resonance of the entropy fluctuations, it is crucial
that the initial value $T_{0}$ be smaller than $T_{\textrm{eq},\psi}$.

\subsection{Growth of the entropy mode}

In the two previous subsections we have seen that, 
in the absence of back-reaction effects, the tachyonic
instability of the entropy mode shuts off once $T > \Teqpsi$.
For the purpose of an order of magnitude estimate for the 
growth, we can take the Floquet exponent $\mu_F$ to be
constant with 
\be \label{Floquet}
\mu_F \, = \, \left(\frac{2\hat{h}_0^{-1/2}}{\alpha'}\right)^{1/2}
= \left( 16 \pi^3\gs\right)^{1/4} M\ .
\ee
Denoting the time when the instability starts with $t = 0$
and the time when it shuts off by $t = t_f$ we have
\be \label{entropygrowth}
\delta s(t) \, = \, e^{\mu_F t} \delta s(0) \ ,
\ee
where $\delta s(0)$ is the initial amplitude of the entropy
fluctutation. The final value of the entropy mode consequently is
\be \label{entropyfinal}
\delta s(t_f) \, = \, e^{\mu_F t_f} \delta s(0) \ .
\ee

Since in the region of instability the growth rates
of the tachyon and that of the entropy mode are the 
same, see (\ref{eq:solT}), we have 
\be
\Teqpsi \, = e^{\mu_F t_f} T_0 \, ,
\ee
and thus 
\be
e^{\mu_F t_f} \, = \, \frac{\delta s(t_f)}{\delta s(0)}\, =\,
\frac{\Teqpsi}{T_0} \ .
\ee

\subsection{Induced growth of the curvature fluctuation}

Having calculated the resonant growth of the entropy mode, we can now
insert the result into the master equation (\ref{eq:changeinR}) for
the induced growth of the comoving curvature perturbation $\R$. On
large scales, and in the region of the tachyonic resonance 
where $|\dot{\psi}|>|\dot{T}|$, equation (\ref{eq:changeinR}) becomes
\bea
\dot{\R}  &\approx&  -\frac{2 H}{\dot{\psi}^2}\,V_T \,\delta s
\nonumber \\
&\approx& \frac{2 H}{\dot{\psi}^2} 2 \hat{h}_0^{-1/2} \alpha'^{-1}\,T\,\delta s
\ .
\eea
Inserting the solutions (\ref{eq:solT}) and (\ref{entropygrowth}) for $T$ and 
$\delta s$, respectively, we find after integration 
%
%%RB \Teqpsi replaced by T(t_f) for more general applicability.
%
\be
\delta\R \, \approx \,  
\frac{H}{\dot{\psi}^2}  \mu_F T(t_f) \,
\delta s(t_f)\ .\label{Scurvature}
\ee
In the absence of back-reaction, $T(t_f) = \Teqpsi$.
Since $\Teqpsi$ is just where the tachyon and $\psi$ velocities are
equal in magnitude, we obtain
\bea 
\delta\R  &\approx& \frac{H}{\mu_F}
\frac{\delta s(0)}{T_0} \nonumber\\
&\approx&\sqrt{\frac{4\pi}{3}}\sqrt{\frac{T_3}{\mpl^4}}
\frac{(\mpl \sqrt{\alp})}{\hat{h}_0^{1/4}}\frac{\delta s(0)}{T_0 }\ .
\label{seccond}\eea
For the parameters from \cite{KKLMMT} cited in section
\ref{subsec:examples} and used throughout this paper, the resulting
amplitude is of the order
\be \label{result}
\delta\R (t_f) \, \approx \, 10^{-5} \,
\frac{\delta s(0)}{T_0} \ .
\ee
We expect that the initial value of the tachyon and the
entropy mode are given by the same quantum fluctuation amplitude
calculated in (\ref{amp}). More specifically, $\delta s(0)$ is
set by $\delta T$ at the wavelength in question, which is given by
equation (\ref{Tatk}).  We then normalize $\delta s(0)$ appropriately for
a power spectrum by multiplying it by $k^{3/2}$ (since our final goal
is calculating the power spectrum of $\R$), and we see that 
$\delta s(0)$ is approximately the same as the quasi-homeogeneous tachyon
value $T_0$ obtained in equation (\ref{Tamp}). 

However, if one were to follow the refined argument made above
concerning the $k$-dependence of this background $T_{0}(k)$, the
wavenumber $k$ at which one measures $T_{0}(k)$ should be larger than
the one considered for $\delta s(0)$. In this way, we can be sure that
the resulting $T_{0}$ is quasi-homogeneous at the scale in question, and
we actually have $T_{0}\leq\delta s(0)$, which only enhances the effect.
However, since we try to set a lower limit, we will be conservative and 
treat $T_{0}$ and $\delta s(0)$ roughly as equal.

Hence, we conclude from (\ref{result})
that, in the absence of back-reaction effects, 
the amplitude of the curvature fluctuations induced by the
entropy modes is comparable to the amplitude of the primordial
linear adiabatic fluctuations.  
Note that our result (\ref{result}) is independent of the specific
value of the intial quasi-homogeneous tachyon amplitude $T_0$. The reason
for this is that the smaller $T_0$ is, the smaller the initial value of
the entropy mode, but the longer the tachyonic instability lasts.

Another useful way to see our 
result is to consider the ratio of these secondary curvature perturbations
to the primary perturbations ($\mathscr{R}=\delta\rho/\rho$):
\eq{Rratio}{
\frac{\delta\mathscr{R}}{\mathscr{R}} = 
\frac{1}{6\pi} \left(\frac{v}{4\pi \gs}\right)^{1/4} \left(\frac{\mpl}{M}
\right)^2\left(\frac{\mu}{\phis}\right)^5 \frac{\delta s(0)}{T_0}\ .}
For the parameter values we use, we find 
$\delta\mathscr{R}/\mathscr{R}\approx 3.9\delta s(0)/T_0$, which implies that
the secondary anisotropies are actually larger than the primary anisotropies.

Let us now consider a different set of parameter values. Compared
to the values from \cite{KKLMMT} used in the bulk of the paper, we can 
rescale $\gs\to x\gs$ and $\mathcal{M}\to \mathcal{M}/x$ without changing 
the warp factor (note that this is a discrete choice because flux is 
quantized and depends on other compactification parameters, as well).  
Working through the details, it is not hard to find that
\eq{eq:gsrescale}{\left(\frac{\delta\rho}{\rho}\right)\to
\left(\frac{\delta\rho}{\rho}\right) x^{-2}\ ,\ \ 
\left(\frac{\delta\mathscr{R}}{\mathscr{R}}\right)\to
\left(\frac{\delta\mathscr{R}}{\mathscr{R}}\right)x^{3/2}\ .
}
In addition, by changing the compactification volume (again, this would
require adjusting other microphysical parameters), we can rescale
$(\alp\mpl^2)\to (\alp\mpl^2)y$.  This is slightly more subtle, because 
this process also rescales the warp factor $\hat h_0$ and the deformation
parameter $\epsilon$ \cite{Giddings:2005ff}.  In the end, we find 
\eq{eq:alprescale}{\left(\frac{\delta\rho}{\rho}\right)\to
\left(\frac{\delta\rho}{\rho}\right) y^{3}\ ,\ \ 
\left(\frac{\delta\mathscr{R}}{\mathscr{R}}\right)\to
\left(\frac{\delta\mathscr{R}}{\mathscr{R}}\right)y^{-11/3}\ .
}
Between the two of these rescalings, it is easy to see that we can maintain
the COBE normalization for the primary anisotropies, while the secondary
anisotropies would increase in amplitude (assuming for convenience
that the primary anisotropies are still calculated using slow-roll physics
at $\phis$).  In other words, we can easily
find parameter values where our results present an even sharper problem.

We have thus established our main result, namely that, %%RB small addition!!
at least in the absence of back-reaction, there are
parameter values in the brane inflation model we have considered
for which the secondary fluctuations are larger than the primary
ones. In order to agree with observations, the model parameters
will thus have to be normalized to the data using the secondary
fluctuations rather than the primary ones. This will lead to different
values of the model parameters which are consistent with the
data.

%%RB New subsection!!

\subsection{Back-reaction effects} \label{s:backreaction}

Although the quasi-homogeneous value of the tachyon field on the
infrared scales relevant to our study is very small, the dispersion
of the tachyon field on microphysical scales is large. Using the
vacuum values of the small-scale tachyon fluctuations it can easily be
shown (see e.g. \cite{hep-th/0106179}) that, at the onset
of the tachyonic instability, the small-scale
disperson $\sigma$ of the tachyon field is of the order $m$, where $m$ is the
effective mass of the tachyon field before the waterfall point is
reached. We denote this initial field dispersion by $\sigma(0)$. 
Its value is much larger than the value $T_{0}$ of the
quasi-homogeneous tachyon field at the onset of the resonance. Due to
the tachyonic resonance, the dispersion $\sigma$ grows exponentially.
After a time interval $t_s$ which is given by $m^{-1}$, the dispersion
has grown to a field value corresponding to the location $\eta$
of the minimum of the tachyon potential. The time $t_s$ is called
the spinodal decomposition time.\footnote{This subsection was added
after very useful discussions with Jim Cline and N. Barnaby.}

Let us model the tachyon potential by the standard potential of a
waterfall field in hybrid inflation
\be \label{hybrid}
V(T) \, = \, \frac{\lambda}{4} \left( T^2 - \eta^2 \right)^2 \, 
\ee
where $\lambda$ is the coupling constant for tachyon field interactions.
Fitting $\lambda$ and $\eta$ to our potential (\ref{eq:pot-reh}), 
we find that $\eta$ is of the order $M^2 / m$. Thus, the spinodal
decomposition time is of the order of
\be
t_s \, \approx \, m^{-1} \ln\left(\frac{\eta}{\sigma(0)}\right) \, .
\ee
This time must be compared with the time $t_f$ when the tachyonic
growth for long wavelength fluctuations stops. This time is given
by
\be
t_f \, \approx \, m^{-1} \ln\left(\frac{\Teqpsi}{T_0}\right) \, .
\ee

Since $T_{0}$ is generically much smaller than $m$, the spinodal 
decomposition time is generically shorter than $t_f$. Thus,
the tachyon field becomes nonlinear on microphysical scales before
the time $t_f$ is reached. If $T_{0}$ is given by 
(\ref{amp}), then the difference between $t_f$ and $t_s$ is only
by a logarithmic factor. However, if $T_{0}$ is exponentially
suppressed by $e^{-3N_{k}/2}$ due to the red-shifting of the long wavelength
tachyon fluctuations during the period of slow-roll inflation,
$N_k$ being the number of e-foldings
of inflation between when the scale $k$ under
consideration exits the Hubble radius and the onset of reheating,
then the ratio $t_f / t_s$ is of the order $N_k$.

The onset of non-linearity on microphysical scales does not in itself
shut off the tachyonic growth on cosmological scales, in the same way
that the gravitational collapse of structures on stellar scales in our
universe has not shut off the linear growth of perturbations on scales
relevant to the cosmic microwave background. 

However, once the tachyon dispersion $\sigma$ approaches the minimum of
the potential, nonlinear effects in the tachyon field equation become
important and generate a positive contribution to the mass term in
the tachyon potential. The magnitude of this back-reaction
effect depends quite sensitively on the form of the potential. 
Working in the context of the above toy model (\ref{hybrid}), we
can make use of the Hartree approximation to estimate the contribution
$\delta m^2_{eff}$ to the effective square mass, and find that
it is of the order
\be
\delta m^2_{\mathrm{eff}} \, \approx \, \lambda \sigma^2\ ,
\ee
which dominates over the negative contribution to the square mass as
soon as
\be
\lambda \sigma^2 \, > \, m^2 \, .
\ee
In the case of the potential (\ref{hybrid}), this happens on a time scale
$t_s$, and greatly suppresses the efficiency of the tachyonic growth
of the entropy fluctuations. In a follow-up study, we plan to study these
back-reaction effects in an actual brane inflation model. 

%%AF New subsection!!

\subsection{Toy model avoiding back-reaction}

It is possible that some models of brane inflation allow the tachyon to be
light during the last 60 e-foldings of inflation, which means that the entropy
mode amplification could continue uninterrupted by back-reaction effects.
We will now discuss a toy model in which the brane and antibrane are both
located at the tip of the deformed conifold, as recently discussed by 
\cite{Pajer:2008uy} (following work by \cite{DeWolfe:2007hd}),
and point out parameter values
in which entropy modes could become important.  

As discussed in \cite{Pajer:2008uy}, nonperturbative corrections to the 
D-brane action can generate a potential for the angular motion of the brane
on the deformed conifold, even in an approximation in which the warp factor
is independent of the angular directions.  In some cases, \cite{Pajer:2008uy}
found that this potential can support an adequate number of e-foldings
of slow-roll inflation; in that case, the potential takes the form
\eq{pajerV}{V\approx 2\Lambda \left(1-\frac{1}{16d^4}\psi^4\right)\ ,}
near the top of the potential.  We now use $\psi$ to denote the angular 
position of the brane, which starts near $\psi=0$, while the antibrane sits
at $\psi=\pi\mpl$ in our model at the antipodal point of the throat's 
tip.\footnote{We use the variable $\psi$ to remind the
reader that this field is the inflaton in a region of approximately constant
warp factor.}  In this model, we consider the nonperturbative
potential to dominate over the Coulomb interaction between the brane and 
antibrane.  Then COBE normalization requires $\Lambda^{1/4}\sim 10^{-3}d$
with $d\lesssim \mpl$.

We can now ask how long 
the tachyon might be light during this type of slow-roll
inflation.  If we combine equations (\ref{eq:ds6warped},\ref{eq:hat-h0}),
we see that the proper radius of the $S^3$ at the tip of the deformed 
conifold throat is, up to factors of order unity, 
$\sqrt{g_s\M\alp}$. For the sample
values given in section \ref{subsec:examples} (and other commonly taken
string parameters), the radius is therefore essentially $\sqrt{\alp}$.
Thus, we see that it is likely that the brane/antibrane tachyon is no more
massive than the warped string scale whenever the brane is on the tip of 
the deformed conifold.  In fact, since the brane and antibrane are
not at antipodal points of the tip during all of inflation, the tachyon 
is likely to be substantially lighter than the warped string scale during 
most of inflation. 

Furthermore, as discussed in \cite{FMM}, the supergravity description of the
warped throat is reasonable as long as the Hubble parameter $H$ is less than
about the warped string scale.\footnote{In \cite{Pajer:2008uy}, it was
assumed that the potential should be less than the warped string scale to
avoid violating the supergravity approximation.  However, since the 
potential energy is due to the interaction between the D3 brane and a D7 brane
elsewhere in the throat, it is not concentrated at the tip of the throat.
What is important is that the 10D potential density be less than the 10D 
string scale, which is possible due to the length of the throat.}
Therefore, if $\Lambda$ and hence $H$ are 
tuned to be large ($H\lesssim \hat h_0^{-1/4}/\sqrt{\alp}$), 
it is reasonable that the tachyon is light compared to the
Hubble scale during much of slow-roll inflation.  In fact, inflation may 
end by hitting the waterfall point rather than violating the slow-roll
conditions. After the waterfall point, we estimate, as before,
the evolution of the tachyon by ignoring its coupling to the inflaton.
While this overestimates the rate of entropy mode growth, it underestimates the
length of time over which the entropy modes can grow.  In fact, while the
Hubble expansion is still important in the background tachyon evolution (that
is, the background tachyon evolution is over-damped), the short
wavelength tachyon fluctuations are also over-damped.  In essence, the 
tachyon fails to roll until the inflaton has moved past the waterfall point.

In that case, super-Hubble fluctuations of the tachyon, and therefore $T_0$,
will be unsuppressed.  This fact means that the long wavelength tachyon
fluctuations at the end of inflation will be just as large as the short
wavelength modes, so back-reaction from the short wavelength modes will not
turn on before the long wavelength entropy modes can amplify the 
curvature perturbation.  In fact, we can immediately estimate the curvature
perturbation.  If we repeat our analysis, the key results 
(\ref{Scurvature}) and the first line of (\ref{seccond}) are unchanged.
In addition, $\mu_F$ is unchanged, still given by the warped string scale.
Therefore, 
\eq{deltaRtoy}{\delta\R \approx \frac{H}{\mu_F} \frac{\delta s(0)}{T_0}
\ ,}
which can be up to order unity.  Therefore, the potential (\ref{pajerV}) 
with appropriate parameter values
provides an explicit example of a brane inflation where entropy
perturbations can affect the normalization of cosmic perturbations to
observation.

%%AF NEW PARAGRAPH
We should note that this model does not have a parametric separation 
between the Hubble and warped string scales as we have presented it.  
Therefore, in order to trust the approximate action we have used for the
D-brane tachyon, which is valid for $\dot\psi,\dot T\lesssim \hat h_0^{-1/2}/
\alpha'$ and $T\lesssim \hat h_0^{-1/4}/\sqrt{\alpha'}$, we need to 
tune the model a bit more.  In particular, if the waterfall point is still
within the slow-roll regime, we can satisfy these constraints
and $T_0<\Teqpsi$ when $\sqrt{\epsilon}\mpl H\lesssim \hat h_0^{-1/2}/\alp$
($\epsilon$ the usual slow-roll parameter).  On the other hand, 
even when $\dot\psi,\dot T$ and $T$ are above the warped string scale, it
is possible that the entropy mode still grows tachyonically at the beginning
of reheating.  To determine whether or not that happens, it is necessary
to analyze the dynamics of the complete tachyonic action as given, for
example, in \cite{hep-th/0411222}.  While that calculation is beyond the
scope of this paper, our results provide considerable motivation for it in
future work.

\section{Discussion and Conclusions}

We have studied the development of entropy fluctuations in brane
inflation models of KKLMMT \cite{KKLMMT} type. We have
shown that the tachyonic instability at the end of slow-roll
inflation leads to an exponential growth of the entropy mode
associated with the tachyon. In turn, the entropy
fluctuations lead to an extra contribution to the curvature
fluctuations. For the parameter
values used in \cite{KKLMMT}, we find that, 
in the absence of back-reaction, the curvature
fluctuations induced by the entropy mode are comparable to 
the primordial curvature fluctuations. For different 
parameter values, we find that the secondary fluctuations may be 
considerably larger than the primary ones.\footnote{A similar conclusion was
recently found in \cite{0704.0212} in the context of
another string-inspired inflationary model, namely ``Roulette'' 
inflation \cite{hep-th/0612197}.} 

However, we have also seen that back-reaction may cut off the
resonant growth of the entropy modes before these have had a chance
to become important. In a simple hybrid inflation model back-reaction
effects indeed will truncate the resonance. Whether this will happen
in any given brane inflation model will require further study.

Our result shows that the dominant source of
curvature fluctuations in brane inflation models
of KKLMMT type may be not the primordial inflaton fluctuations,
but rather the entropy fluctuations occuring at the end of inflation, 
see \cite{Lyth3}.  In this sense, the mechanism is
a realization of the ``curvaton'' \cite{Mollerach:1989hu,gr-qc/9502002,
astro-ph/9610219,hep-ph/0110096,hep-ph/0110002,hep-ph/0109214} scenario.
In order for such a model to produce the observed magnitude
of density fluctuations, the value of the inflaton mass
scale $M$ must not be fixed by (\ref{primcond}), but
by demanding that the amplitude from (\ref{seccond})
yields the observed value. 

Our work is closely related to that of \cite{astro-ph/0601481,Barnaby}
which also discussed entropy modes arising during the early phase of tachyon 
condensation. The methods used are slightly different. We have
introduced an effectively homogeneous background produced by 
long wavelength fluctuations and reduced the subsequent analysis to
a first-order perturbative calculation, whereas 
\cite{astro-ph/0601481,Barnaby} 
assumed that the tachyon averages to zero and studied the generation of
entropy fluctuations using techniques of second order perturbation theory.

Finally, we give a few caveats and directions for further research.
The most important issue is to clarify the strength of back-reaction
effects in the brane inflation model at hand. We should also note that, 
if the time scale for the tachyon evolution is longer than the Hubble
time, there will be no exponential instability for entropy
fluctuations because the friction term in (\ref{eq:eofm-T}) cannot be
neglected (recall that the entropy mode and tachyon are essentially identical
early in reheating).  However, this seems unlikely because the tachyon 
evolution is dominated by the warped string scale for 
$\psi\lesssim \psis/\sqrt{2}$, which we estimate occurs in less
than a Hubble time.

In addition, the Coulombic potential we used is very much a toy model
for brane inflation.
We should also address how the results we have found 
might appear in more complete inflection-point inflation
models of brane inflation \cite{arXiv:0705.3837,arXiv:0706.0360} (see also
\cite{Allahverdi:2006iq,Allahverdi:2006wt,Allahverdi:2006we,Allahverdi:2007wh} 
for a similar form of inflection-point inflation in the MSSM).  
There are two main issues that would need to be addressed.  First,
the inflaton will not follow slow-roll behavior at the onset of tachyon
condensation.  Instead, the D3 brane will oscillate around the bottom of 
the throat, perhaps just exiting from a stage of DBI inflation 
\cite{hep-th/0310221}.  The second issue to address, which
is perhaps more difficult, regards the initial condition for the tachyon at 
the waterfall point.  In our model, we have used the known behavior of a 
scalar field during inflation; however, at the inflection point, the tachyon
would be so massive that it should be integrated out.  The tachyon would
be deflected from $T=0$ instead during the early stages of inflaton 
oscillation or during DBI inflation.  

We can make a few comments going beyond our toy model, however.  If the 
tachyon still has a canonical kinetic term (which may be modified somewhat
for DBI inflation), the tachyon and entropy mode growth is essentially
unchanged from our toy model.  Therefore, the key results 
(\ref{Scurvature}) and the first line of (\ref{seccond}) are also unchanged.
In addition, $\mu_F$ is unchanged, while $H$ should be no smaller than the
value we used.  In that respect, we believe that our results give a lower
bound on the contribution of entropy modes to curvature perturbations.

In our analysis we have neglected cosmic string production at the
end of the phase of tachyon condensation   
\cite{hep-th/0204074,hep-th/0412095}. Once
the tachyon approaches the minimum of its potential, string
production and interaction dominates the energy transfer 
\cite{hep-ph/0012142,hep-th/0106179}.
%However, since the growth of the entropy modes which we are
%studying takes place in the initial period of tachyon growth,
%%RB rest of the paragraph changed!!
Since the characteristic length scale of string production
is much smaller than the Hubble radius, we do not expect
this process to effect the long wavelength entropy modes
studied here. 
%created in the initial phase.

We view our work as an initial step at exploring the vast terrain
of entropy fluctuation modes in string-inspired inflationary
universe models. As shown here, these modes have the potential
to rule out large classes of models and to change the parameter
values in others. As we comment above, some of our key results may very
well apply to more realistic models of inflection point inflation, so it 
will be important to resolve this issue for those potentials, as well.

\section*{acknowledgments} 
 
We would like to acknowledge extremely stimulating conversations 
with N. Barnaby, J. Cline, and R. Danos.  
We also thank N. Barnaby and S. Panda for pointing out references.

The work of RB and AF is supported by an NSERC Discovery Grant. RB
is also supported in part by the Canada Research Chairs program and
by funds from a FQRNT Team Grant. 
AF is supported in part by the Institute for Particle Physics and the Perimeter
Institute. LL is supported by the DAAD.

%\bibliography{entcurve}

\end{document}